\documentclass[aps,twocolumn,preprintnumbers,raggedbottom,prb,amsmath,showpacs,nobalancelastpage]{revtex4}

\usepackage{epsfig,psfrag}
\usepackage{dcolumn}     
\usepackage{bm}          
\usepackage{amssymb}          
\usepackage{times}

\begin{document}

\title{Non-Markovian dynamics of double quantum dot charge qubits due
to acoustic phonons}

\author{M.~Thorwart$^1$, J.~Eckel$^1$, and E.~R.~Mucciolo$^{2,3}$}

\affiliation{$^1$Institut f\"ur Theoretische Physik,
Heinrich-Heine-Universit\"at D\"usseldorf, D-40225 D\"usseldorf,
Germany \\ $^2$ Department of Physics, University of Central Florida,
Box 162385, Orlando, FL 32816-2385, USA \\ $^3$ Departamento de
F\a'{\i}sica, Pontif\a'{\i}cia Universidade Cat\'olica do Rio de
Janeiro, C.P. 37801, 22452-970 Rio de Janeiro, Brazil}

\date{\today}

\begin{abstract}
We investigate the dynamics of a double quantum dot charge qubit which
is coupled to piezoelectric acoustic phonons, appropriate for GaAs
heterostructures. At low temperatures, the phonon bath induces a
non-Markovian dynamical behavior of the oscillations between the two
charge states of the double quantum dot. Upon applying the numerically
exact quasiadiabatic propagator path-integral scheme, the reduced
density matrix of the charge qubit is calculated, thereby avoiding the
Born-Markov approximation. This allows a systematic study of the
dependence of the $Q$-factor on the lattice temperature, on the size
of the quantum dots, as well as on the interdot coupling. We calculate
the $Q$-factor for a recently realized experimental setup and find
that it is two orders of magnitudes larger than the measured value,
indicating that the decoherence due to phonons is a subordinate
mechanism.
\end{abstract}

\pacs{03.67.Lx, 03.65.Yz, 73.63.Kv, 63.20.Kr}

\maketitle

\section{Introduction}

Various candidates for realizing building blocks of quantum
information processors with nanoscale solid state structures have been
proposed and also partially realized in ground-breaking
experiments. An important class of proposals consists of using the
charge degree of freedom in semiconducting double quantum dots (DQDs)
\cite{wielrmp,brandes} to realize a quantum mechanical two-state
system or quantum bit
(qubit).\cite{Blick,Tanamoto,Fedichkin,Brandes02} Thereby, the logical
states $|0\rangle$ and $|1\rangle$ are given by the low-energy charge
states of the DQD with one excess electron either on the left or the
right dot. The transition between these states occurs via tunneling
between the two dots. A significant advantage of the charge qubit is
its direct controllability via external voltage sources. In recent
experiments, the coherent manipulation of charge states of DQDs has
been achieved.\cite{Hayashi03,Petta04,Emiroglu05} Thereby, the control
over the bias and the gate voltages permits to reliably tune the DQD
in the Coulomb blockade regime to the required transition line $(1,0)
\leftrightarrow (0,1)$. The large charging energies suppress leakage
to energetically higher lying many-electron states.
 
 A central impediment for coherent quantum information processes is
 decoherence and dissipation, see Ref.\ \onlinecite{Thorwart02} for a
 particular example. Since charge qubits are rather easily addressable
 from outside, they are, in turn, also rather fragile for various
 decoherence mechanisms from the environment. In order to achieve a
 thorough understanding of the role of the various decoherence
 mechanisms, one has to rely on realistic and precise model
 calculations which allow to sort out the different
 contributions. Among them, decoherence due to acoustic phonons is one
 possible candidate. This mechanism has been studied previously upon
 using approaches relying on the Born-Markov
 approximation.\cite{Eduardo04,brandes,Brandes02,Stavrou05} The
 decoherence rates obtained in these studies were one or two orders of
 magnitude smaller than those determined experimentally for GaAs
 DQDs.\cite{Hayashi03,Petta04} However, no exact treatment was
 available to validate the Born-Markov approximation in the DQD
 context. This is particularly interesting in view of realistic DQD
 designs since geometrical form factors tailor a rather peculiar
 spectral density $J(\omega)$ of the phonon environment. For
 piezoelectric phonons, it has a high-frequency tail decaying only
 algebraically with \cite{Eduardo04} $J(\omega) \propto \omega^{-1}$
 while it is superohmic at low frequencies with $J(\omega) \propto
 \omega^{3}$, with a crossover at frequencies in the regime of the
 tunneling amplitude of the DQD. As it has also been pointed out in
 Ref.\ \onlinecite{Eduardo04}, the use of the Born-Markov
 approximation is appropriate only at small tunneling
 amplitudes. However, is is expected to become increasingly unreliable
 at DQD with larger interdot tunneling amplitudes. The motivation to
 find exact reference solutions for this problems also stems from the
 fact that the Born-Markov approximation has led to the conclusion
 that other mechanisms of decoherence, as for instance, background
 charge fluctuations and electromagnetic noise from voltage
 fluctuations, would be the dominating source of decoherence while
 phonon decoherence should be a negligible contribution. In view of
 further design optimization for real devices, exact results are
 clearly desirable.
 
 In this paper, we study phonon decoherence using the same model as in
 Ref.\ \onlinecite{Eduardo04} as a basis for the numerically exact
 iterative quasiadiabatic propagator path integral (QUAPI)
 scheme. \cite{QUAPI,Tho98,Tho00} In particular, we avoid the
 Born-Markov approximation which turns out to be problematic at larger
 tunneling amplitudes $\Delta$ of the DQD. The decaying 
  oscillations of DQD charge states allows us to extract
 the quality (or Q-) factor. We find that $Q$ decreases with
 increasing tunneling amplitudes $\Delta$ for small $\Delta$. In fact,
 the solution in this regime is accurately described by the
 weak-coupling result for a superohmic environment.\cite{Wei} At large
 $\Delta$, the $Q$-factor increases with increasing $\Delta$ and the
 non-Markovian corrections become noticeable changing even the order
 of magnitude of $Q$. In between these two regimes, a crossover
 occurs which is related to the geometry of the DQD, see below. In
 particular, our numerical precise method allows to obtain results for
 the DQD device recently realized in
 experiments.\cite{Hayashi03,Petta04} We find that the calculated
 value lies approximately two orders of magnitude above the
 experimentally measured value. Indeed, this suggests that the phonon
 decoherence mechanism is subordinate and other mechanisms dominate
 decoherence in DQD charge states which have been realized until now. 
In comparison to the results from
 the Born-Markov approximation, we find that the exact $Q$-factor is in
 general larger, indicating that the former overestimates phonon
 decoherence.
 
\section{Model for charge qubit and phonon bath}

In order to solely investigate the role of the phonons in the
decoherence processes, we start with the simplifying assumption that
the DQD is isolated from the leads. The total Hamiltonian is given
by\cite{Brandes02,Wei}
\begin{equation}
H=H_S + H_{B} + H_{SB} \, .
\end{equation}
Here, $H_S$ is the Hamiltonian of the DQD which is modeled in the
basis of the two localized charge states $|0\rangle \equiv |L\rangle
\equiv (1,0)$ and $|1\rangle \equiv |R\rangle \equiv (0,1)$ as a
symmetric two-level-system in pseudo-spin notation ($\sigma_i$ are the
Pauli matrices) as
\begin{equation}
H_S = \hbar \Delta \sigma_x \, .
\end{equation} 
In this simple model, we have assumed that the external gate voltage
is tuned such that the two charge states are energetically nearly
degenerate and that the excess electron can tunnel between the two
dots with the tunneling amplitude $\Delta$. A possible energy bias
between the two states could easily be included in the model but is
not considered in this work.
 
The Hamiltonian for the phonon bath is, as usual, given by
\begin{equation}
H_B=\hbar \sum_{\mathbf q} \omega_{\mathbf q} b^\dagger_{\mathbf q} 
b_{\mathbf q} \, , 
\end{equation} 
with the phonon dispersion relation $\omega_{\mathbf q}$. 

\begin{figure}
\begin{center}
\epsfig{figure=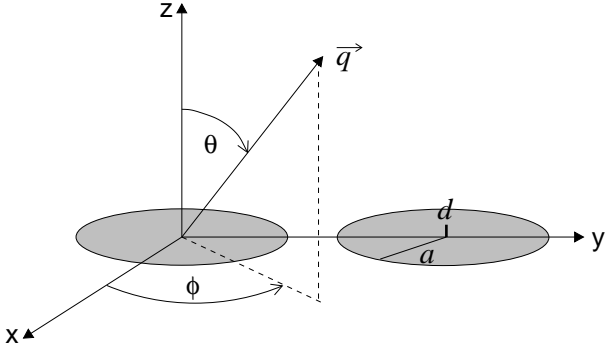,width=80mm,angle=0,keepaspectratio=true}
\caption{Sketch of the geometry of the double quantum dot charge 
qubit and the various angles of the phonon propagation. 
 \label{fig.geometry}}
 \end{center}
\end{figure}

\begin{figure}
\begin{center}
\epsfig{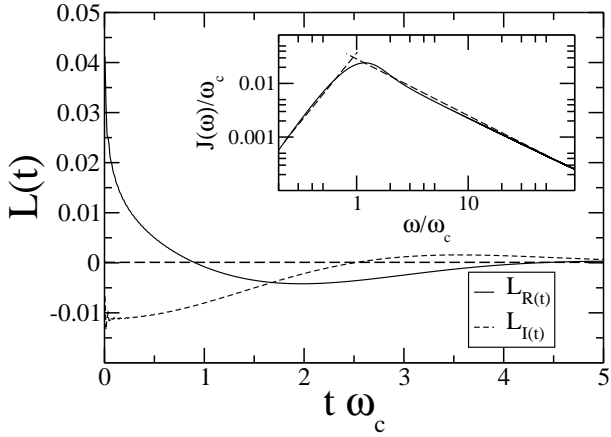}
\caption{The bath autocorrelation function (response function)
$L(t)=L_R(t) + i L_I(t)$ for the spectral density $J(\omega)$ (inset)
of the phonon bath for the case of GaAs with $g_{\rm ph}=0.05$, $s=5
\times 10^{3}$ m/s, a dot radius of $a=60$ nm, and interdot distance
$d=180$ nm. Temperature is $T=15$ mK. The dashed lines in the inset
mark the superohmic limit $J(\omega \rightarrow 0) =\alpha_{\rm SO}
\omega_c^{-2} \omega^3$ at low frequencies and the high-frequency
limit $J(\omega \rightarrow \infty) \propto 1/\omega$.
 \label{fig.specdens}}
 \end{center}
\end{figure}

The interaction part of the Hamiltonian is determined by the
electron-phonon interaction and reads
\cite{brandes99,brandes,Brandes02,Eduardo04}
\begin{equation}
H_{SB} = \hbar \sum_{\mathbf q} (\alpha_{\mathbf q}^{L} N_L + 
\alpha_{\mathbf q}^{R} N_R ) (b^\dagger_{\mathbf q}+b_{-\mathbf q}) \,
.
\end{equation} 
Here, $N_\xi$ is the number of the excess electrons on the left/right
($\xi=$L/R) dot and $\alpha_{\mathbf q}^{\xi} = \lambda_{\mathbf q}
e^{-i {\mathbf q} \cdot {\mathbf R}_{\xi}} F_\xi ({\mathbf q})$ where
${\mathbf R}_{L}=0$ and ${\mathbf R}_{L}=d {\mathbf e}_y$ are the
position vectors of the two dots. As in Ref.\ \onlinecite{Eduardo04},
we have assumed that the center of the left dot is located at the
origin of the coordinate system while the center of the right dot is
placed on the $y$-axis with distance $d$, see Fig.\  
\ref{fig.geometry} for a sketch. The two dots are assumed to
have an equal radius of $a$ and to be confined to the $xy$-plane in
the underlying 2DEG. The coupling constants $\lambda_{\mathbf q}$
which depend on material parameters and on the wave vector ${\mathbf
q}$ will be specified below. To take into account a realistic dot
geometry, we associate a form factor
\begin{equation}
F_\xi ({\mathbf q}) = \int d^3r \, n_\xi({\mathbf r}) e^{-i{\mathbf q}
\cdot {\mathbf r}}
\end{equation}
to each dot, where $ n_\xi({\mathbf r})$ is the charge density
distribution of the dot. In the following, the two dots are assumed to
have identical form factors. With this the coupling Hamiltonian
simplifies to \cite{brandes99,Eduardo04}
\begin{equation}
H_{SB} = \frac{\hbar}{2} \sigma_z \sum_{\mathbf q} g_{\mathbf q} 
(b^\dagger_{\mathbf q}+b_{-\mathbf q}) \, , 
\end{equation}
with $g_{\mathbf q}=\lambda_{\mathbf q} F({\mathbf q}) (1-e^{-i
{\mathbf q} \cdot d{\mathbf e}_y})$. Notice that the phonons can
propagate in all three dimensions and that the
electron-phonon-coupling is not isotropic.\cite{Eduardo04} Having
specified the coupling, we can now introduce the spectral density of
the phonon bath which reads
\begin{equation}
J(\omega) = \sum_{\mathbf q} |g_{\mathbf q}|^2 \delta(\omega - 
\omega_{\mathbf q}) \, . 
\end{equation}
As in Ref.\ \onlinecite{Eduardo04}, we specialize to linear acoustic
phonons with velocity $s$ ($s\approx 5 \times 10^{3}$ m/s for GaAs)
and linear dispersion $\omega_{\mathbf q} = s |{\mathbf
q}|$. Moreover, we only consider coupling to longitudinal
piezoelectric phonons and neglect the contribution of the deformation
potential which is justifiable at temperatures below 10 K for bulk
GaAs material.\cite{Eduardo04,note} This yields the coupling constant
$|\lambda_{\mathbf q}|^2 = g_{\rm ph} \pi^2 s^2 / (V |{\mathbf q}|)$,
where $g_{\rm ph}$ is the dimensionless piezoelectric constant
($g_{\rm ph} \approx 0.05$ for GaAs) and $V$ being the volume of the
unit cell. Assuming that the charge density distribution on the dot is
Gaussian in the $xy$-plane and localized in the $z$-direction, one
finds for the spectral density of the bath \cite{Eduardo04}
\begin{eqnarray}
J(\omega) & = &  g_{\rm ph} \omega \int_0^{\pi/2} d\theta 
\sin \theta e^{-\omega^2a^2\sin^2 \theta / s^2} \nonumber \\
& & \times \left[ 1- J_0\left(\frac{\omega d }{s} \sin \theta 
\right)\right] \, . 
\label{specdens}
\end{eqnarray}
The spectral density is shown in the inset of Fig.\ \ref{fig.specdens}
for the case of GaAs and a dot radius of $a=60$ nm. The low-frequency
behavior is superohmic, i.e., $J(\omega \rightarrow 0) \approx
\alpha_{\rm SO} \omega_c^{-2}\omega^3$ with $\alpha_{\rm SO}= g_{\rm
ph} d^2 /(6a^2)$, while the high-frequency tail falls off
algebraically like $J(\omega \rightarrow \infty) \propto
1/\omega$. The crossover between these two limiting regimes occurs on
a frequency scale $\omega_c=s/a\equiv \tau_c^{-1}$. For GaAs and a
typical dot radius of $a=60$ nm, we obtain
$\omega_c=83$ GHz corresponding to an energy of $55\,\mu$eV. As we
will see below, the typical tunneling amplitudes $\Delta$ are
comparable to this energy scale. This indicates that the Markov
approximation could become problematic in this transition region. The
cut-off frequency $\omega_c$ corresponds to the inverse of the time
scale $\tau_c$ of the bath autocorrelation function which reads
\cite{Wei}
\begin{eqnarray}
L(t) & = & L_R(t)+iL_I(t) \nonumber \\
& = & \frac{1}{\pi} \int_0^\infty d\omega J(\omega) \left[ \coth
\frac{\hbar \omega \beta}{2} \cos \omega t - i \sin \omega t
\right]. \label{response}
\end{eqnarray} 
The autocorrelation function $L(t)$ is shown in Fig.\
\ref{fig.specdens} for $T=15$ mK for $a=60$ nm and $d=180$
nm.\cite{Eduardo04,Jeong01} 
The algebraic decay at high frequencies determines the short time
behavior of $L(t)$; notice the finite slope of $L_R(t)$ at $t=0$, in
contrast to zero slope of the usual exponential and Drude-shaped
cutoff functions.\cite{Wei} The superohmic low frequency behavior of
$J(\omega)$ yields a fast decay of $L(t)$ at long times. The often
used Born-Markov approximation corresponds to replacing the strongly
peaked real part $L_R(t)$ by a $\delta$-function with the
corresponding weight while the imaginary part $L_I(t)$ is often
neglected. Moreover, we note that the dimensionless damping constant
given by the prefactor $\alpha_{\rm SO}$ of the low-frequency
superohmic limit assumes the value $\alpha_{\rm SO}=0.075$ for the
parameters quoted above. Thus, our results at small $\Delta$ can be
compared to those from a weak-coupling approach in terms of real-time
path-integrals for the spin-boson model with a superohmic
environment.\cite{Wei}
\section{The quasiadiabatic path-integral propagator (QUAPI)} 
The dynamics of the qubit is characterized by the time evolution of
the reduced density matrix $\rho(t)$ which is obtained after tracing
out the bath degrees of freedom, i.e.,
\begin{eqnarray}
\rho(t) & = & {\rm tr}_B U(t,0) W(0) U^{-1} (t,0)\, , \nonumber \\
U(t,0) & = & {\cal T} \exp \left\{ -\frac{i}{\hbar} \int_0^t 
dt' H \right\} \, . \label{rhored}
\end{eqnarray}
Here, $U(t,0)$ denotes the propagator of the full system plus bath and
${\cal T}$ denotes the time-ordering operator. $W(0)$ is the full
density operator at initial time set at $t=0$. We assume standard
factorizing initial conditions,\cite{Wei} i.e., $W(0) \propto \rho(0)
\exp [-H_B/(k_B T)]$, where the bath is at thermal equilibrium at
temperature $T$ and the system is prepared according to $\rho(0)$.
Throughout this work, we always start from the qubit state $\rho(0) =
|L\rangle \langle L |$.

The technique which we use to calculate $\rho(t)$ is the well
established iterative tensor multiplication algorithm derived for the
quasiadiabatic propagator path integral (QUAPI).\cite{QUAPI} It is a
numerically exact algorithm, as also, for instance, the real-time
quantum Monte Carlo method is.\cite{Egger} It has been successfully
tested and adopted in various problems of open quantum systems, with
and without external driving.\cite{QUAPI,Tho98,Tho00} For details of
this algorithm, we refer to previous work \cite{QUAPI,Tho98,Tho00} and
here only address the ingredients which are important to our
calculations.
 
The algorithm relies on the symmetric Trotter splitting of the
short-time propagator $U(t_{k+1},t_k)$ into system ($H_S$) and bath
($H_B$) parts on a time slice of length $\Delta t$. The bath dynamics
can be solved exactly yielding a Feynman-Vernon-type influence
functional \cite{Wei} while the system is propagated exactly by
solving the Schr\"odinger equation. The neglect of higher order terms
of the full propagator causes an error of the order of $\Delta t ^3$.  
The Trotter splitting allows to calculate an approximate value of the 
true result for the observable of interest,  with an error depending on $\Delta t$. 
As shown in Ref.\ \onlinecite{Fye86}, this error vanishes proportional 
to $\Delta t^2$ as $\Delta t \rightarrow 0$. Thus, by extrapolation 
of the results for  $\Delta t \rightarrow 0$, the  Trotter error 
can be eliminated completely (the details of 
this procedure will be discussed elsewhere\cite{Eckel05}). 

The interaction of the system with the bath induces internal
transitions in both and creates intercorrelations between them
(memory); the latter are described by the autocorrelation function
$L(t)$ given in Eq.\ (\ref{response}). For the phonon bath, these
correlations decay on a time scale given by the correlation time
$\tau_c$. This motivates us to neglect such long-time correlations
beyond a memory time $\tau_{\rm mem}$ and to break up the influence
kernels into pieces of total length $\tau_{\rm mem}= K \, \Delta t$,
where $K$ denotes the number of time steps over which the memory is
fully taken into account.

The two strategies mentioned above are countercurrent: For the Trotter
splitting, a small time step $\Delta t$ is desirable, thus decreasing
$\tau_{\rm mem}$ for a fixed $K$. On the other hand, a large
$\tau_{\rm mem}$ is wanted in order to take a long memory range into
account. Nevertheless, converged results can be obtained in the following way: 
(i) We choose $\tau_{\rm mem}$ such that we include all the relevant parts of 
the correlation function $L(t)$. Quantitatively, this can be done by requiring 
$L_R(\tau_{\rm mem})/L_R(0) \le 10^{-3}$ in the asymptotic regime. 
(ii) We choose $K$ such that the resulting $\Delta t=\tau_{\rm mem} / K$ 
is small enough to ensure that we are in the regime which allows 
extrapolation for $\Delta t \rightarrow 0$ (see above). This procedure allows to 
completely eliminate the Trotter error.  We note that the choice of $K$ is limited to a 
maximum of $K=12$ for reasonable numerical practicability on a personal computer. 

\section{The $Q$-factor of coherent charge oscillations}
\begin{figure}
\begin{center}
\epsfig{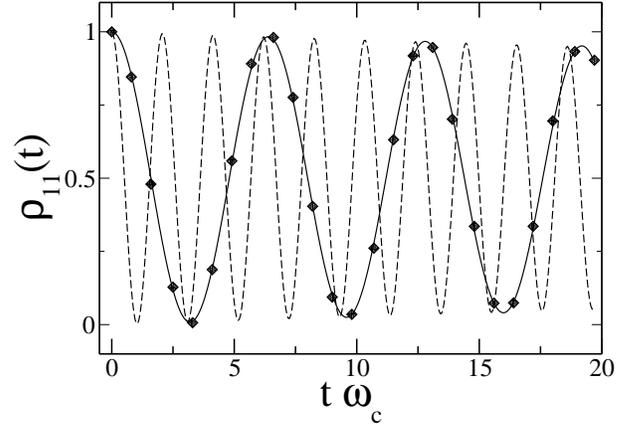}
\caption{Time evolution of the charge population of 
the left dot for the case $\Delta=27 \mu$eV and $T=15$ mK. Shown are the 
exact results obtained from QUAPI ($\blacklozenge$) and the result of a 
 fit of an exponentially decaying cosine
to $\rho_{11}(t)$ (solid line).  
The dashed line indicates the result of a weak-coupling (Born-Markov) 
approximation using Eqs.\ (\ref{freq}) and (\ref{gam}) for the same parameters. 
Deviations from the exact results are striking.  
 The remaining parameters are the same as
in Fig.\ \ref{fig.specdens}.
 \label{fig.QUAPI}}
 \end{center}
\end{figure}

We have adjusted the QUAPI algorithm to the phonon spectral density
(\ref{specdens}) and determine the charge population $\rho_{11}(t)$ of
the left dot as a function of time for the initial condition $\rho_{11}(0)=1$. 
 A typical example of the resulting damped oscillatory behavior 
is shown in Fig.\ \ref{fig.QUAPI} where the exact QUAPI results are depicted 
by the symbols ($\blacklozenge$) for the case $\Delta = 27 \mu$eV, $T=15$ mK and $K=12$.  
To extract the 
damping rate $\gamma$ and the oscillation frequency $\Omega$, we fit
an exponentially decaying cosine to the numerical data. The result of the 
fit is also shown in Fig.\ \ref{fig.QUAPI} as solid line. 
The fit yields $\Omega_{\rm QUAPI} = 
0.98 \omega_c$ for the oscillation frequency and $\gamma_{\rm QUAPI} = 5.27 
\times 10^{-3} \omega_c$ for the decay constant. 
  
\begin{figure}
\begin{center}
\epsfig{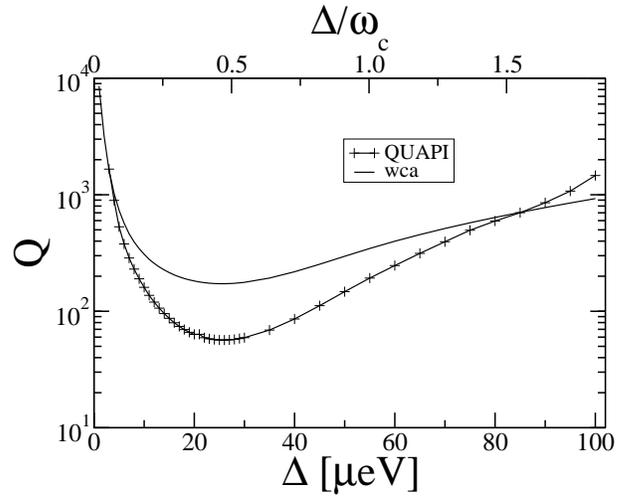}
\caption{The quality factor $Q$ of the coherent charge oscillations as
a function of the tunneling amplitude $\Delta$ in natural units (lower
scale) and scaled with respect to $\omega_c$ (upper scale). The data
shown are obtained with QUAPI ($+$) using the phonon spectral density
(Eq.\ (\ref{specdens}) and with the weak-coupling approximation (wca)
of Eqs.\ (\ref{freq}) and (\ref{gam}). The parameters are the same as
in Fig.\ \ref{fig.specdens}.
\label{fig.delta}}
 \end{center}
\end{figure}

 The ratio of $\gamma$ and $\Omega$ fixes the quality factor according
 to the convention $Q=\Omega / (\pi \gamma)$. Evaluating $Q$ allows us
 to investigate the dependence of the coherence of the charge
 oscillations on various experimental parameters. Figure
 \ref{fig.delta} shows the $Q$-factor as a function of the tunneling
 amplitude $\Delta$ obtained with the numerically exact QUAPI
 algorithm ($+$). For illustration, we have used parameters from the
 experimental setup of Ref.\ \onlinecite{Jeong01}. Since our model
 contains realistic assumptions about the geometry of the DQD and the
 materials characteristics, the predictions of the QUAPI calculations
 in this case can be considered quite accurate. Two regimes exist: (i)
 At small $\Delta$, $Q$ decreases with increasing $\Delta$ (see below)
 while (ii) at large $\Delta$, $Q$ increases again with increasing
 $\Delta$. The crossover in between occurs when the energy splitting
 between the qubit states coincides with the maximum of the spectral
 density $J(\omega)$, i.e., when $2 \Delta = \omega_c$. Then, the
 damping is most efficient and thus the rate $\gamma$ is
 maximal. Notice that realistic values for the tunneling amplitudes
 $\Delta$ fall in the range of the bath correlation frequency scale
 $\omega_c$ (see upper axis of Fig.\ \ref{fig.delta}, where $\Delta$
 is scaled with respect to $\omega_c$).
  
  Since the coupling to the phonons is rather weak (cf.\ the value of
  $\alpha_{\rm SO}=0.075$ for the superohmic coupling constant at low
  frequencies), it is tempting to compare our exact results to simple,
  approximate analytical results obtained from real-time path-integral
  formalism in the regime of weak-coupling.$\cite{Wei}$
  However, we remark that one needs to have in mind that the algebraic
  decay of the spectral density at large frequencies is by itself not
  taken into account while deriving the approximate result. Thus, only
  the superohmic low-frequency part can be expected to yield a
  reasonable agreement. The weak-coupling results are given
  by\cite{Wei} 
  \begin{equation}\label{freq}
  \Omega=2 \Delta [1-2 \, {\rm Re } \, u(2 i\Delta)]^{1/2}
  \end{equation}
  and
  \begin{equation}\label{gam}
  \gamma=\frac{1}{4} J (2 \Delta) \coth \frac{\hbar \Delta}{k_B T} \, ,
  \end{equation}
  where 
  \begin{equation}
  u(z)=\frac{1}{2} \int_0^\infty d\omega
  \frac{J(\omega)}{\omega^2+z^2}\left( \coth \frac{\hbar \omega}{2 k_B T}
  -1\right) \, .
  \end{equation}
With this, we can compute the time evolution of charge population of the 
left dot in the weak-coupling approximation which is shown in 
Fig.\  \ref{fig.QUAPI} (dashed line) for the same parameters as above. 
The deviation from the exact result for the chosen  parameters due to 
non-Markovian effects is striking and also illustrated by 
the numerical values $\Omega=3.0 \omega_c$ and $\gamma = 5.6 \times 10^{-3} 
 \omega_c$. 

The resulting $Q$-factor is shown in Fig.\ \ref{fig.delta} (solid line,
``wca''). Since the results in Eqs.\ (\ref{freq}) and (\ref{gam}) are
obtained by linearization with respect to the damping constant $g_{\rm
ph}$, they are equivalent to a Born-Markovian approximation. We find
noticeable deviations for intermediate and large values of 
$\Delta$. From Eq.\ (\ref{freq}) it follows that
the oscillation frequency $\Omega$ is renormalized by the term
involving $u(2 i\Delta)$ stemming from one-phonon interblip
correlations in the self-energy.\cite{Wei} We find that at small
$\Delta$ these corrections are negligible, while they increasingly
become important at larger $\Delta$, pointing to a non-Markovian
dynamics in this regime. Note that the qualitative behavior of $Q$
versus $\Delta$ is similar to that found in
Ref. \onlinecite{Eduardo04} within the Born-Markovian master equation
approach. However, the absolute numbers disagree.
 
\begin{figure}
\begin{center}
\epsfig{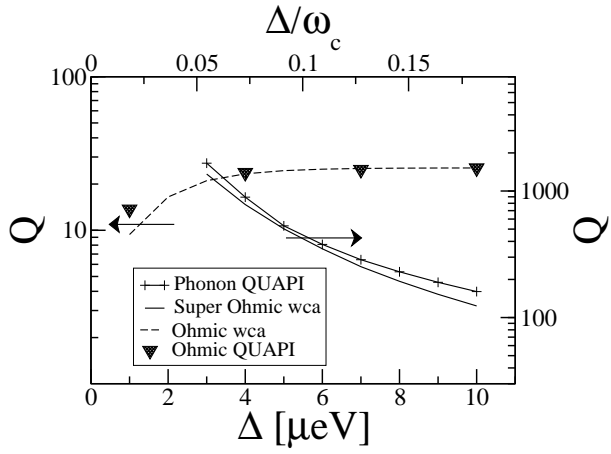}
\caption{Zoom into the small-$\Delta$ region. Shown are the QUAPI
results for the phonon ($+$) and the pure Ohmic ($\blacktriangledown$)
spectral density. In addition, they are compared with the approximate
analytical, weak-coupling results of Eqs.\ (\ref{freq}) and
(\ref{gam}) for the superohmic (solid line) and the ohmic (dashed
line) cases. The parameters are: (i) Ohmic case, $\alpha=0.05$ and
(ii); superohmic case, $\alpha_{\rm SO}=0.075$. For all cases, $T=15$
mK.
 \label{fig.ohm_supohm}}
 \end{center}
\end{figure}

 To understand the results at small tunneling amplitudes, we zoom into
 the small-$\Delta$ region and show the QUAPI results in Fig.\
 \ref{fig.ohm_supohm} ($+$, right ordinate scale) together with the
 results from the weak-coupling approximation (solid line) obtained
 from Eqs.\ (\ref{freq}) and (\ref{gam}). Therefore, we have used the pure
 superohmic spectral density with an exponential cutoff with frequency
 $\Omega_c\gg \Delta$, i.e, $J(\omega)=\alpha_{\rm SO}
 \omega_c^{-2}\omega^3 \exp(-\omega/\Omega_c)$ with $\alpha_{\rm
 SO}=0.075$ and $\Omega_c=10 \omega_c$. We find a good agreement
 between both results in the limit of small $\Delta$, thus justifying
 that the dynamics at small $\Delta$ is Markovian. Nevertheless, note
 that even in this frequency range deviations between the exact and
 the approximate solution do appear. As a further check, we also
 compare the results of both the numerically exact and the analytical
 approach for $Q$ for a pure Ohmic environmental spectral density
 $J(\omega)=\alpha \omega \exp(-\omega/\omega_c)$ for $\alpha=0.05$. 
 Both the QUAPI ($\blacktriangledown$) and the
 approximative weak-coupling solution (dashed line) coincide. The
 $Q$-factor decreases monotonically for decreasing $\Delta$ since the
 Ohmic low-frequency modes are not suppressed strongly enough in
 comparison with the superohmic case. Although the dynamics is
 Markovian for small $\Delta$, non-Markovian corrections appear to be
 relevant at intermediate values of $\Delta$. Because realistic values
 for the tunneling splitting $\Delta$ are also of the order of
 $\omega_c$, the role of the large frequency tail $\propto 1/\omega$
 still is not negligible and cannot be appropriately taken into
 account within a Born-Markovian master equation.
  
\begin{figure}
\begin{center}
\epsfig{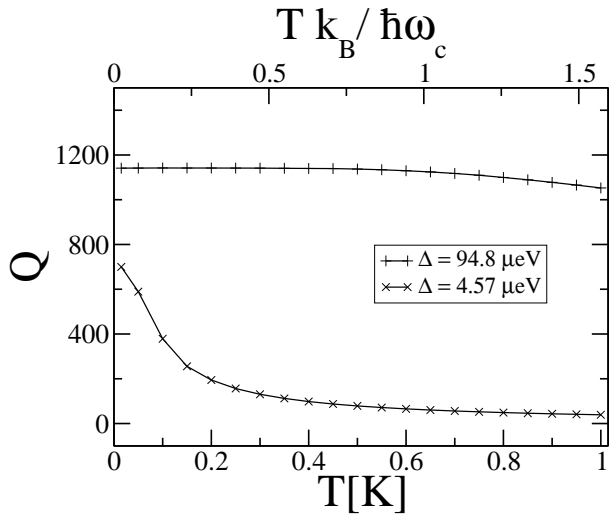}
\caption{$Q$-factor as a function of temperature for two different
tunneling amplitudes $\Delta$. The remaining parameters are the same
as in Fig.\ \ref{fig.specdens}.
 \label{fig.temp}}
 \end{center}
\end{figure}

\begin{figure}
\begin{center}
\epsfig{figure=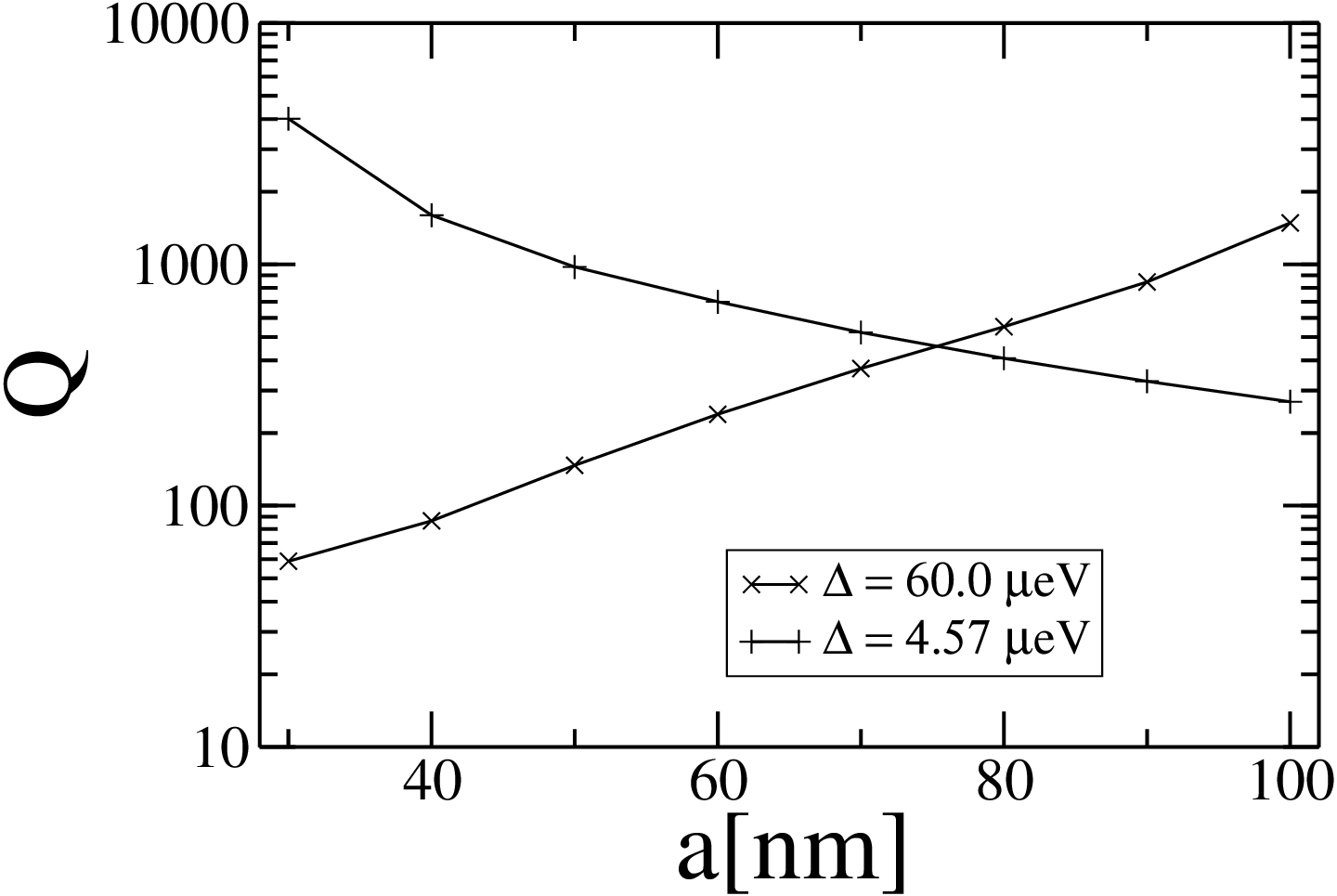,width=80mm,angle=0,keepaspectratio=true}
\caption{$Q$-factor as a function of the dot radius $a$ for two
different tunneling amplitudes $\Delta$ and for a fixed ratio $d/a=3$.
The remaining parameters are the same as in Fig.\ \ref{fig.specdens}.
 \label{fig.size}}
 \end{center}
\end{figure}

The sensitivity of the $Q$-factor on the lattice temperature $T$ of
the GaAs host is shown in Fig.\ \ref{fig.temp} for a small and a large
tunneling amplitude $\Delta$. Note that for the latter the corresponding 
Born-Markov results have been shown not to be reliable due to 
noticeable non-Markovian corrections which can be observed in our data. 
 Another parameter which is adjustable
in the DQD design is the dot radius $a$. The dependence of $Q$ on $a$
is depicted in Fig.\ \ref{fig.size} for a large and a small tunneling
amplitude for a fixed ratio $d/a=3$. We find that $Q$ decreases
monotonically with increasing $a$ for small value of $\Delta$, while
it increases with increasing $a$ for large value of $\Delta$,
qualitatively similar to the findings in Ref.\ \onlinecite{Eduardo04}. Note, however, 
the sizeable differences in the absolute numbers which are due to the non-Markovian
corrections. 

Finally, we address the recent experiment of Hayashi {\em et
al.\/}.\cite{Hayashi03} These authors implemented a GaAs DQD in the
bias-pulsing mode. During the pulse, the tunneling amplitude is
constant. The bias pulse is ramped up on a time scale of 100 ps. The
tunneling amplitude has been estimated as $\Delta=5$ $\mu$eV. The
lattice temperature was $T=20$ mK, while the effective dot radius and
the interdot distance have been estimated to be $a \approx 50$ nm and
$d\approx 225$ nm, respectively. For this combination of parameters,
the QUAPI method yields $Q=352$ which has to be compared with the
experimental value of $Q=3$. 
On the other hand, the weak-coupling solution of Eqs.\ (\ref{freq})
and (\ref{gam}) yields $Q=539$. Two important conclusions can be drawn
from these numbers. First, the Born-Markov approximation
underestimates decoherence in these systems and non-Markovian
corrections are quite noticeable. Second, our exact result differs
from the experimental value by a factor of roughly $100$, indicating
that the role of the phonons for the decoherence is indeed of minor
importance to understand the reported experimental 
results. Other sources of decoherence like voltage fluctuations
from the electromagnetic environment or background charge fluctuations
exist and have to be taken into account for an accurate description
for so far realized charge-based DQD qubits. 

However, our result go beyond this point: Our predicted 
exact values of the $Q$-factors represent a fundamental upper limit on the 
possibility to improve coherence of DQD charge qubits since phonon 
decoherence represents an intrinsic decoherence mechanism which 
can hardly be overcome. This has to be noticed also in view 
of the DiVincenzo criteria \cite{DiVincenzo} for a possible 
realization of a quantum computer.

\section{Conclusions}

To summarize, we have obtained numerically exact results for the
$Q$-factor of the decaying charge oscillations in a double quantum dot
upon using the real-time quasiadiabatic propagator path integral
(QUAPI). Realistic assumptions on the form of the environmental phonon
spectral density entered in our model via geometrical form factors and
materials characteristics. No fitting parameters of any sort were
utilized. We have investigated the quality ($Q$-) factor as a function
of the tunneling splitting and have compared our results with those
obtained from a weak-coupling approximation within an analytical
approach in terms of real-time path-integrals. We find that the regime
of small tunneling amplitudes is appropriately covered by the
Markovian description. However, at larger (but still realistic) values
of the tunneling amplitude, non-Markovian corrections appears and are
relevant. Moreover, we have determined the temperature dependence of 
 $Q$-factor as well as its dependence on the dot radius. 
From a comparison with the result obtained in an
experimental realization of a double quantum dot charge qubit, we find
that the theory predicts a $Q$-factor two orders of magnitudes larger
than the measured value. This leads to the conclusion that phonon
decoherence is a subordinate mechanism in GaAs quantum dots realized 
with present state-of-the-art technology. Clearly, other 
forms of coupling to the external environment prevail for those 
already realized devices. However, our results also represent a fundamental 
upper limit to the coherence of DQD charge qubits which can hardly be 
overcome due to its intrinsic nature. 

It would be
interesting to extend the QUAPI calculation to Si-based charge qubits,
which have recently been implemented.\cite{Emiroglu05} Since Si has
inversion symmetry, piezoelectric phonons do not occur and one needs
to include deformation potential phonons in the calculations. The
decoherence induced by the electron-phonon coupling is likely to be
smaller than in GaAs. However, for the deformation potential case,
$\lambda_{\bf q} \sim \sqrt{|{\bf q}|}$. Thus, if on one hand we
expect a strong superohmic behavior at low frequencies, on the other
hand, a very slow decay of the spectral function at high frequencies
should occur. That may provide an even stronger non-Markovian dynamics
and have implications for other implementations, such as Si:P
donor-based charge qubits.\cite{barrett,australians,belita}

\acknowledgements{We thank Harold Baranger, Stephan Weiss, Reinhold Egger and 
Frank Wilhelm for discussions.}

\newpage

\begin{thebibliography}{00}
\bibitem{note} As shown in Ref. \onlinecite{Stavrou05}, deformation
potential phonons are only important for ultra small DQD
setups. Optical phonons are too high in energy ( $\sim$ 36 meV) and
can be safely neglected.
\end{thebibliography}
\end{document}